
\magnification=1200
\pretolerance=10000
\hsize=4.5 in
\hoffset=1.0 cm
 \overfullrule = 0pt
\line{ }
\vskip 0.5 truecm
\rightline {DFTUZ 93.08, May 1993}
\vskip 2. truecm
\centerline{\bf FERMIONIC EFFECTIVE ACTION AND THE PHASE STRUCTURE}
\centerline{\bf OF NON COMPACT QUANTUM ELECTRODYNAMICS IN 2+1 DIMENSIONS}
\vskip 2 truecm
\centerline { V.~Azcoiti, X.Q. Luo and C.E. Piedrafita  }
\vskip 0.15 truecm
\centerline {\it Departamento de F\'\i sica Te\'orica, Facultad de
Ciencias, Universidad de Zaragoza,}
\centerline {\it 50009 Zaragoza (Spain)}
\vskip 0.5 truecm
\centerline { G. Di Carlo and A.F. Grillo }
\vskip 0.15 truecm
\centerline {\it Istituto Nazionale di Fisica Nucleare, Laboratori
Nazionali di Frascati,}
\centerline {\it P.O.B. 13 - Frascati (Italy). }
\vskip 0.5 truecm
\centerline { A. Galante}
\vskip 0.15 truecm
\centerline {\it L.N.F - I.N.F.N. and Dipartimento di Fisica, Universita'
dell' Aquila, L' Aquila 67100, (Italy)}
\vskip 3 truecm
\centerline {ABSTRACT}
We study the phase diagram of non compact 
$QED_3$ using the microcanonical fermionic average method described 
elsewhere. We present evidence for a continuous phase transition line 
in the $\beta, N$ plane, extending down to 
arbitrarily small flavour number $N$.

\vfill\eject

\par
\noindent

In this Letter we continue a systematic study of Lattice Abelian models with
dynamical fermions [1,2,3] that we are making through the use of the
microcanonical average method [1] for dealing with fermionic lattice 
simulations.

Earlier work was related to four dimensional compact  [1], 
and non compact [2,3] models. In particular the non
compact abelian model is interesting, being a candidate for a strongly
interacting continuum theory: in this case we have presented detailed 
analysis of the phase structure, both in $(\beta,m)$, 
including $m=0$ [2] and $(\beta, N$) [3] planes.

Lower dimensional models have a particular interest on theoretical 
grounds.
In $1+1$ dimensions, quantum electrodynamics at zero mass ( {\it i.e.} the
Schwinger model) is analytically solvable, confining and asymptotically
free; in this case lattice results can be directly related to exact 
(continuum) ones.

The $2+1$ dimensional Abelian model shares some features with Schwinger's:
it confines static charges and is asymptotically free,
so its understanding should be relevant 
to more physical theories in four dimensions. Although not solvable, 
the model is super renormalizable. 
It is also potentially interesting in relation to models
of high $T_c$ superconductivity [4].

In this paper we describe our study of the phase structure of
non compact $QED_3$ in the $(\beta, N)$ plane. We find 
a critical line that consists of two segments, a first order line 
in the large-$N$ and a second order one in the small-$N$ region. 
Our analysis is based on the studies of two quantities, 
the effective fermionic action and the chiral condensate. 
An intriguing feature that emerges from these studies is 
that the phase diagrams
obtained using the two quantities differ in the low-$N$ region. The analysis
of the fermionic action suggests that the second order critical line 
terminates on the $N=0$ axis at finite coupling, $\beta_c=0.49$, implying
the existence of two phases even in the quenched theory. The behavior of 
the chiral condensate, however, shows no restoration of chiral 
symmetry for any finite $\beta$, in the investigated range,
implying that the nonanaliticities occuring at $\beta_c=0.49$ are not related
to the chiral symmetry breaking, in agreement with the conclusion of previous
studies of $QED_3$ [5,6,7] as well as with theoretical expectations.

We mention that extensive amount of work on the subject has been
done in the continuum formulation of the model.
This has been reported in refs.[8-11]. 
Also, simulations of the compact variant of $QED_3$, with the 
use of the method described below, have been presented in [12].

The method we use, 
is based on the introduction of an effective fermionic
action $S^F_{eff}(E,N,m)$ [1-3]. An advantage of this method is 
that it allows simulations to be performed exactly in the chiral limit.
The effective fermionic action, which is a function of the
pure gauge energy $E$, fermion mass $m$ and number of flavours $N$,
is defined through,

$$  e^{-S_{eff}^F(E,m,N)} = $$
$$ {\int [dA_{\mu}(x)] (\det
\Delta(m, A_{\mu}(x)))^{N/2}
\delta({1\over 2} \sum_{x,\mu < \nu} F_{\mu \nu}^2(x) - 3VE)
\over
\int [dA_{\mu}(x)]
\delta({1\over 2} \sum_{x,\mu < \nu} F_{\mu \nu}^2(x) - 3VE)}
\eqno(1)$$

\noindent
where $\Delta(m, A_\mu(x))$
is the fermionic matrix (we use 4-components staggered fermions) and $E$ is
the normalized pure gauge energy. The denominator in (1) is the density of
states at fixed energy

$$ N(E)= C_G E^{V - {3\over 2}} \eqno(2)$$

\noindent
with $C_G$ being an unimportant (divergent) constant and $V$ the lattice
volume.

After the definition of the fermionic effective action, 
the partition function of this model can be written as a 
one-dimensional integral

$${\cal Z} = \int dE N(E) e^{-3 \beta V E-S_{eff}^F(E,N,m)} \eqno(3)$$

\noindent
from which we can define an effective full action per unit volume as

$$ \bar S_{eff}(E,\beta,N,m) =
-\ln E + 3 \beta E + \bar S_{eff}^F(E,N,m)  \eqno(4) $$

\noindent
$\bar S^F_{eff}(E,N,m)$
in (4) is the fermionic effective action (1) normalized to the
lattice volume.

Once the effective action, which entirely characterizes the fermionic
contribution, is defined, the qualitative features of the
phase structure of the system can be studied analytically. 
The integrand in (3) is a strongly peaked function of $E$. Since the 
effective full action diverges lineary with the lattice volume $V$, one can 
evaluate the free energy and its derivatives by saddle point.
The mean plaquette energy $<E_p>=E_0(m,\beta,N)$
will be given by the solution of the saddle point equation [2]

$${1 \over {3E}}-\beta -{1\over 3}{\partial\over
\partial E} \bar S^F_{eff}(E,N,m)= 0 \eqno (5) $$

\noindent
satisfying the minimum condition

$${1 \over {E^2}}+{\partial^2\over
\partial E^2} \bar S^F_{eff}(E,N,m) > 0 \eqno (6) $$

By differentiating
equation (5) respect to $\beta$ we get the specific heat

$$ C_{\beta} =  {\partial \over \partial \beta} <E_p>=
 - \{{1\over 3E^2_0(m,\beta,N)}+{1\over3}{\partial^2\over\partial E^2}
\bar S^F_{eff}(E,N,m)\Big|_{E_0(m,\beta,N)} \}^{-1}\eqno(7)$$

The effective action must be continuous even in the thermodynamical limit; 
however its derivatives may be discontinuous in this limit. A discontinuity 
in the first energy derivative implies an analogous 
behaviour for the average energy, hence a first order phase transition.
If the first energy derivative is continuous but some higher order derivative 
is discontinuous, then the average 
energy will be continuous, the specific heat or some of its derivatives being 
discontinuous, {\it i.e.} the 
system will undergo a continuous phase transition [2].

This is not the only way in which a phase transition is generated, however,
since this can be obtained also through a cancellation of the 
two terms of the denominator in (7). 
In order that this happen, the second energy derivative of
the effective action should be negative. Then, for a range of energies 
and $N$ large enough, the denominator will be negative and there will be no 
solution of the saddle point equation, {\it i.e.} there will be a 
range of energies not accessible to the system, indicating again 
a first order  transition.

Finally, if the effective action is non analytic in the flavour number 
$N$ or the fermion mass $m$, this will cause again phase transitions 
in the $(\beta, N)$, $(\beta, m)$ planes.

To summarize, the phase structure of the theory can be entirely described in
terms of the behaviour of the effective action as a function of the
energy and the bare parameters. If it is non analytic in the energy, 
then a phase transition will appear. 
On the other hand, if the effective action is
analytic in $E$, $N$ and $m$, then the only other mechanism 
for producing a phase transition, if any,
will generate a first order transition line ending in a second order point. 
Obviously these mechanisms can coexist, as in $QED_4$ [3].

We stress that this characterization of the phase structure of the
theory (more transparent in the non compact model 
since the density of states is known
analytically), is rather
independent on the details of the numerical evaluation of the effective
action. Also, it does not depend on the evaluation and extrapolation of the
chiral condensate, which, as for past experience, is very delicate 
especially in small lattices and in the three dimensional theories [6]. 
In fact, the above characterization of the phase structure is entirely 
independent from those existing in the literature for these models. 
The phase structure of the four-dimensional theory,
obtained in this way [2] is in very good agreement with the one obtained in 
[13].

We now present our results for the effective action.
For the fermionic effective action we can write down an expansion 
in cumulants [2],

$$-S^F_{eff}(E,N,m)= {N\over 2} <\ln \det \Delta(m,A_{\mu}(x))>_E $$
$$+ {N^2\over 8} \{<(\ln \det \Delta)^2>_E-<\ln \det \Delta>_E^2 \} + ...
\eqno(8)$$

\noindent
where $<O>_E$ means the mean value of the operator $O(A_\mu(x))$ computed
with the probability distribution $[dA_{\mu}(x)]
\delta({1\over 2} \sum_{x,\mu < \nu} F_{\mu \nu}^2(x) - 3VE)/N(E)$.
Expression (8) is a $N$ expansion of the fermionic effective action.

Following the general method described in [1,2], we have done simulations in
$6^3, 10^3, 14^3, 18^3$ and $20^3$ lattices. In Fig. 1 we present the 
Fermionic Effective Action for two massless flavours on an $14^3$ lattice, 
computed using the first two contributions to the cumulant expansion. 
The third one has been found compatible with zero within errors.
The result shown in Fig. 1 is qualitatively completely analogous to the one
found in the four dimensional theory, so we can repeat the same analysis as 
in [2,3]. The important point here is that our data strongly indicate a
{\it continuous} phase transition; in fact the Fermionic Effective Action 
is linear for small energies and clearly not linear 
for larger energies, thus suggesting a discontinuity of the second or higher 
order derivatives in the thermodynamical limit at some critical energy $E_c$. 
The first derivative is continuous, 
as follows from the analysis of the numerical data.

This behaviour is dramatically evident if one plots the Fermionic Effective 
Action minus the fit to its linear part, Fig. 2. Once the critical energy 
$E_c$ is determined by fitting the results reported in Fig.2 with a power law 
function $C(E-E_c)^\rho$, the 
critical coupling is computed from saddle point equations. Notice that at 
these lattice sizes, the results from the saddle point aproximation are 
undistinguishable from the numerical ones.

In Table I we present the critical values of $\beta$ at
various values of $N$. The
transition line continues down to arbitrarily small values of $N$, 
including zero flavour (the quenched theory). The critical values of $\beta$ 
and $E$ in the quenched limit, $\beta_c=0.49(1), E_c=0.68(1)$, 
show no variations with the lattice size, 
in disagreement with expectations for a phase transition with divergent 
correlation length, and might be interpreted as indicating that we are 
indeed observing a transition with finite correlation length. However 
the similarities of these results with those obtained in the four-dimensional 
non compact model [2,3], cast doubts on the previous interpretation. In fact 
our determination of the critical couplings $\beta_c$ in the four-dimensional 
model with two and four dynamical flavours, were in very good agreement 
with those reported in [13], the last obtained in much larger lattices. 
But there is no doubt that the correlation length diverges at the critical 
point of noncompact QED in four dimensions.

Concerning the phase structure at large number of flavours, here the 
discussion in [3] also 
applies: at the critical values $N=6.10, E_c=1.58$ the denominator 
in (6) is zero and for larger values of $N$ it becomes negative in some 
energy interval. In such an energy interval, the saddle point 
equation has no solution, producing a first order
transition line. As described in [3] the continuous transition line 
merges into the
first order one, since the energy at which the effective action becomes 
non analytic falls into the energy interval not
accessible to the system, which widens with the energy. Notice that the 
critical energy obtained from our simulations seems to be independent on 
the flavour number at small $N$, as in the four dimensional case [2,3].
In Fig. 3 we
present the complete ($\beta, N$) phase diagram of the model, at $m=0$.
This phase diagram has been obtained using only the first term in the 
expansion (8).

The only results on the phase structure for the three-dimensional non compact 
case are those in [5,6,7], suggesting a continuous, chiral symmetry 
restoring transition ending at $N \simeq 3-4, \beta=\infty$.
On the other hand, the results reported in [6] show unambiguously that chiral 
symmetry is spontaneously broken in the quenched model for $\beta$ values 
larger than our $\beta_c=0.49$, thus suggesting that the phase transition we 
observe does not restore chiral symmetry, contrary to what happens in the 
four dimensional non compact model. This is not surprising since 
quenched $QED_3$ confines static charges for any finite $\beta$ and there 
are general arguments suggesting that confining forces make the chiral 
symmetric vacuum unstable [14]. In fact our numerical results for the 
chiral condensate in the quenched model support well this scenario. In 
Fig. 4 we plot the inverse logarithm of the chiral condensate against the 
inverse logarithm of the fermion mass 
for several values of $\beta$ in the $14^3$ and  
$18^3$ lattices. This kind of plot was proposed in [13] as a very 
efficient way to get the critical $\beta$ and the value of the 
$\delta$ exponent in a continuous chiral restoring transition. The results 
of Fig. 4 show unambiguously that the "critical $\beta$" obtained in this 
way moves significantly towards  larger values 
when the lattice size changes from $14^3$ to $18^3$, 
indicating that the observation of a "critical $\beta$" 
(which in this plot corresponds to a straight line which 
passes through the origin) is a pure finite size effect, the critical
coupling being pushed towards $\infty$ in the thermodynamical limit.

We try here to discuss on the reliability of our results concerning the 
continuous phase transition line at small $N$. A first criticism 
might be that the method 
we use forces for some reason a phase transition through a (non physical) 
non analiticity of the effective action. We argue that
this is extremely unlikely: on one hand, our determination of the phase 
diagram  of $QED_4$ through the Effective Action is
in very good agreement with those obtained using traditional methods (using 
the behaviour of the chiral condensate); this agreement extends to all 
the physical observables measured [2]. On the other hand we have obtained 
within this approach preliminary results for massless $QED_2$ 
(the Schwinger model), showing a good analytical behaviour of $S_{eff}$, 
{\it i.e.} the absence of phase transitions at finite $\beta$ in the one 
flavour model, with a scaling behaviour in agreement with simple dimensional 
counting.

In Fig. 5 we plot the mean value of the normalized singular part of 
the fermionic 
action in the quenched limit against $\beta$ in a $18^3$ lattice. This 
singular part is defined as 

$$S_F^{sing}(\beta) = {1\over V}<log det\Delta> - a_0 - 
{a_1\over{3\beta}}\eqno(9)$$

\noindent
where $a_0=0.145$, $a_1=-0.256$ are the zero energy intercept and the slope 
in the small energy region of 
the first cumulant contribution to the fermionic effective action (8). The 
existence of two phases is evident in this figure. 
From a numerical point of view, we must note that our critical coupling 
at $N=0$, $\beta_c=0.49$, corresponds to a region where the results 
for the chiral condensate reported in [5,6] show a very rapid change. 
Unfortunately this $\beta$ region has not been explored intensively in [5,6], 
so no definite conclusions can be extracted from their results.

The phase structure of the model in the $(\beta, N)$ plane (Fig. 3) shows 
the existence of two completely separated phases. However, chiral 
symmetry should be spontaneously broken in both phases 
since it is broken for small $N$ including the 
quenched limit. We have also explored if the continuous phase transition line 
is related to the percolation of topological structures associated to the 
lattice regularization. However, 
the monopole [15] and string densities at the critical values of $\beta$ 
are too small to produce percolation of these objects. Therefore we have 
at the moment no compelling evidence for this interpretation. 
We want to notice here that this continuous transition was not observed in the 
simulations of the compact version of this model [12] thus suggesting again 
important qualitative differences between the compact and non-compact 
regularizations, like in the four dimensional case.

Conversely the first order line is clearly produced by pure fermionic 
effects, like in the four dimensional non compact model. We would like 
to remark that our quantitative results for large $N$ could change when 
including all the cumulants in the expansion of the effective action. 
What can be analytically proved is that the effective action is linear 
with $N$ in the large $N$ limit [3] so higher order terms in the 
cumulant expansion conspire between them in order to give this linear 
behaviour. However the main qualitative features like the fact that the 
first order line does not intercept the $\beta =0$ axis [3] remain 
unchanged.

Concerning the possibility to have a chiral restoring phase transition, we 
have not seen evidence for such a transition. However we can not exclude 
a non analiticity of the effective fermionic action as a function of $N$ 
which could originate this transition. Indeed our approach is based 
in an expansion of the fermionic effective action in powers of $N$ and 
the implicit assumption that the convergence radius of this expansion is 
$\infty$.

Several interesting issues emerging out of this study and which are left open 
at this moment are: the physical origin of the continuous 
transition, qualitative differences between the strong and weak coupling 
phases, correlation length and critical exponents associated with the 
continuous transition, possibility to define a 
non superrenormalizable, but renormalizable field 
theory, etc.. 
A more detailed study of these issues is underway.

The numerical simulations quoted above have been done using the Transputer 
Networks of the Theoretical Group of the Frascati National Laboratories, of 
the University of L' Aquila and the Reconfigurable Transputer Network (RTN), 
a 64 Transputers array of the University of Zaragoza.

We  thank A. Kocic for extremely interesting discussions and for 
a critical reading of the manuscript.

This work has been partly supported through a CICYT (Spain) - 
INFN (Italy)
collaboration. 

\vfill
\eject

\line{}
\centerline {\bf REFERENCES}

\vskip 1 truecm
\vskip .1 truecm
\item {1.} 
{V. Azcoiti, G. Di Carlo and A.F. Grillo,
Phys. Rev. Lett. {\bf 65} (1990) 2239;
V. Azcoiti, A. Cruz, G. Di Carlo, A.F. Grillo and A. Vladikas,
Phys. Rev. {\bf D43} (1991) 3487; 
V. Azcoiti et al. "The microcanonical fermionic average method for Monte 
Carlo simulations of lattice gauge theories with dynamical fermions", 
INFN preprint (1992) {\bf LNF-93/004(P)} (1993), 
to appear in  Phys. Rev. {\bf D}.\hfill}

\vskip .1 truecm
\item {2.} {V. Azcoiti, G. Di Carlo and A.F. Grillo, 
Mod. Phys. Lett. {\bf A7} (1992) 3561; 
V. Azcoiti, G. Di Carlo and A.F. Grillo, "A New Approach to
Non Compact Lattice QED with Light Fermions ", 
{\bf DFTUZ 91.34} (1992), to appear in Int. Jour. Mod. Phys. {\bf A}.\hfill}

\vskip .1 truecm
\item {3.} {V. Azcoiti, G. Di Carlo and A.F. Grillo, 
 Phys. Lett. {\bf 305B} (1993) 275.\hfill}

\vskip .1 truecm
\item {4.} {E. Dagotto, E. Fradkin and A. Moreo, 
Phys. Rev. {\bf B38} (1988) 2926.\hfill}

\vskip .1 truecm
\item {5.} {E. Dagotto, J.B. Kogut and A. Kocic, Phys. Rev. Lett. {\bf 62}
(1989) 1083; E. Dagotto, A. Kocic and J.B. Kogut Nucl. Phys. {\bf B334}
(1990) 279.\hfill}

\vskip .1 truecm
\item {6.} {S. Hands and J.B. Kogut, Nucl. Phys. {\bf B335} 
(1990) 455. \hfill}

\vskip .1 truecm
\item {7.} {J.B. Kogut and J.-F. Lagae
Nucl. Phys. {\bf B} (Proc. Suppl.) {\bf 30} (1993) 737. \hfill}

\vskip .1 truecm
\item {8.} {R. Pisarski, 
 Phys. Rev. {\bf D29} (1984) 2423;
 Phys. Rev. {\bf D44} (1991) 1866.\hfill} 

\vskip .1 truecm
\item {9.} {T. Appelquist, M. Bowick, E. Choler and L.C.R. Wijewardhana,
 Phys. Rev. Lett. {\bf 55} (1985) 1715;
T. Appelquist, M. Bowick, D. Karabali and L.C.R. Wijewardhana,
Phys. Rev. {\bf D33} (1986) 3704; 
T. Appelquist, D. Nash and L.C.R. Wijewardhana,
 Phys. Rev. Lett. {\bf 60} (1988) 2575; 
D. Nash, Phys. Rev. Lett. {\bf 62} (1989) 3024.\hfill} 

\vskip .1 truecm
\item {10.} {M.R. Pennington and D. Walsh, 
 Phys. Lett. {\bf 253B} (1991) 246; 
 D.C. Curtis, M.R. Pennington and D. Walsh,
 Phys. Lett. {\bf 295B} (1992) 313.\hfill}

\vskip .1 truecm
\item {11.} {K.I. Kondo and H. Nakatani, 
 Progr. Theor. Phys. {\bf 87} (1992) 193.\hfill}

\vskip .1 truecm
\item {12.} {V. Azcoiti and X.Q. Luo, 
Nucl. Phys. {\bf B} (Proc. Suppl.) {\bf 30} (1993) 741; 
V. Azcoiti and X.Q. Luo {\bf DFTUZ.92/25} (1992). \hfill}

\vskip .1 truecm
\item {13.} {S.J. Hands, A. Kocic, J.B. Kogut, R.L. Renken, 
D.K. Sinclair and K.C. Wang, "Spectroscopy, Equation of State and monopole 
percolation in lattice QED with two flavours",
 {CERN-TH.6609/92} (1992);
A. Kocic, J.B. Kogut and K.C. Wang,
" Monopole Percolation and the Universality Class of the Chiral
Transition in Four flavour non compact Lattice QED."
 ILL-TH-92-17 (1992).\hfill}

\vskip .1 truecm
\item {14.} {R. Brout, F. Englert and J.M. Frere, 
 Nucl. Phys. {\bf B134} (1978) 327;
A. Casher, Phys. Lett. {\bf B83} (1979) 395; 
A. Amer, A. LeYaouanc, L. Oliver, O. Pene and J.C. Raynal, 
Phys. Rev. Lett. {\bf 50} (1983) 87.\hfill}

\vskip .1 truecm
\item {15.} {H.R. Fiebig and R.M. Woloshyn, 
 Phys. Rev. {\bf D42} (1990) 3520.\hfill} 

\vfill
\eject

\line{}
\vskip 1 truecm
\centerline{\bf FIGURE CAPTIONS}
\vskip 1 truecm

\item {1)}{Normalized fermionic effective action as a function of the 
pure gauge energy on a $14^3$ lattice, $m=0.0$ and $N=2$.}

\vskip .1 truecm

\item {2)}{ The order parameter, obtained from the results of Fig. 1}

\vskip .1 truecm

\item {3)}{ Phase diagram on the $(\beta, N)$ plane. }

\vskip .1 truecm

\item {4)}{ Inverse logarithm of the chiral condensate against the inverse 
fermion mass logarithm for several values of $\beta$ in the $14^3$ $(4a)$ and 
$18^3$ $(4b)$ lattices (quenched case).}

\vskip .1 truecm

\item {5)}{ Singular part of the mean value of the effective action normalized 
by the lattice volume $V$ against $\beta$ in a $18^3$ lattice 
(quenched case).}

\vfill
\eject

\line{}
\vskip 1 truecm
\centerline{\bf TABLE CAPTION}
\vskip 1 truecm

\item {I)}{ Critical values of $\beta$ at several values of $N$ on the 
$18^3$ lattice.}

\vfill
\eject

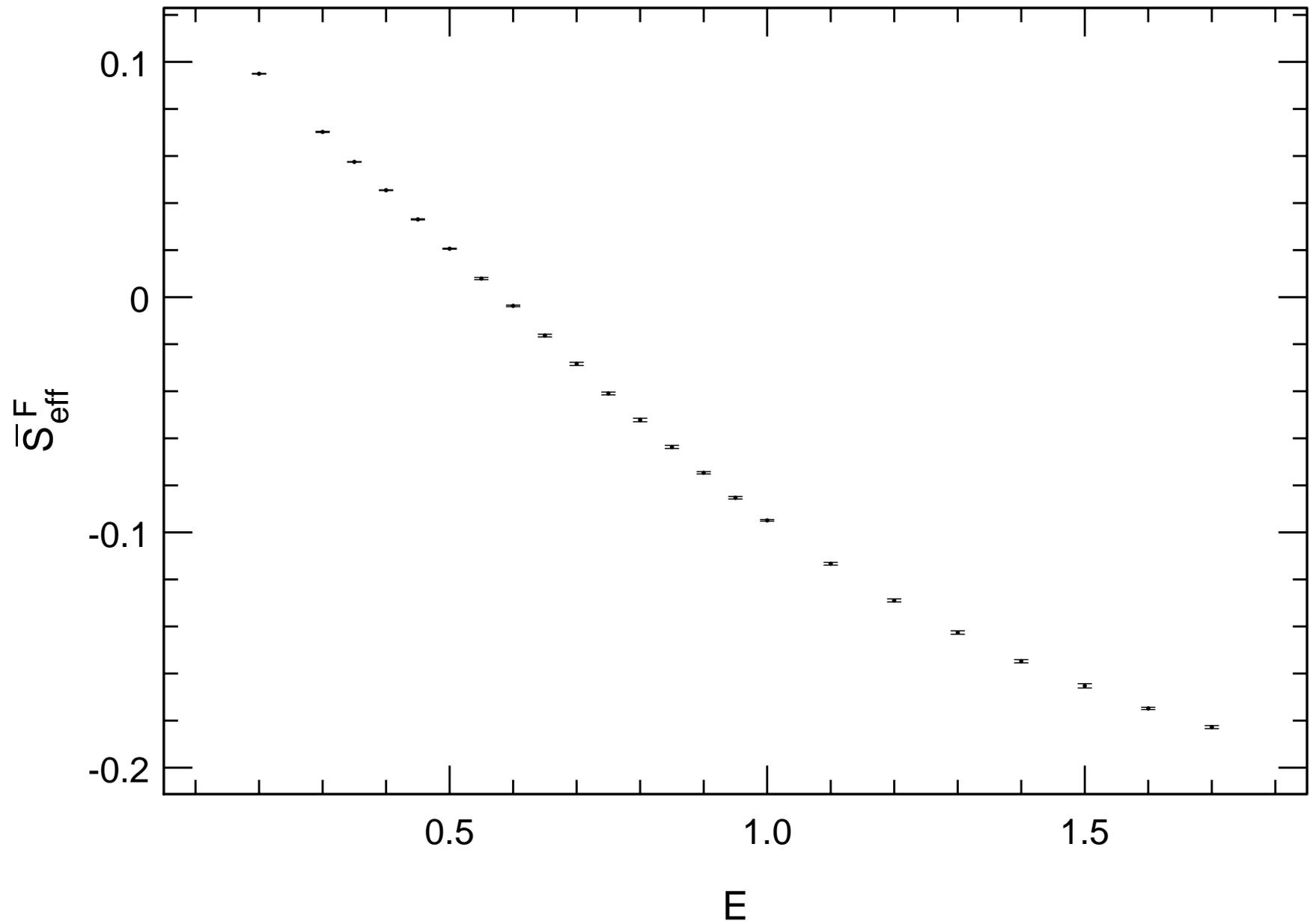
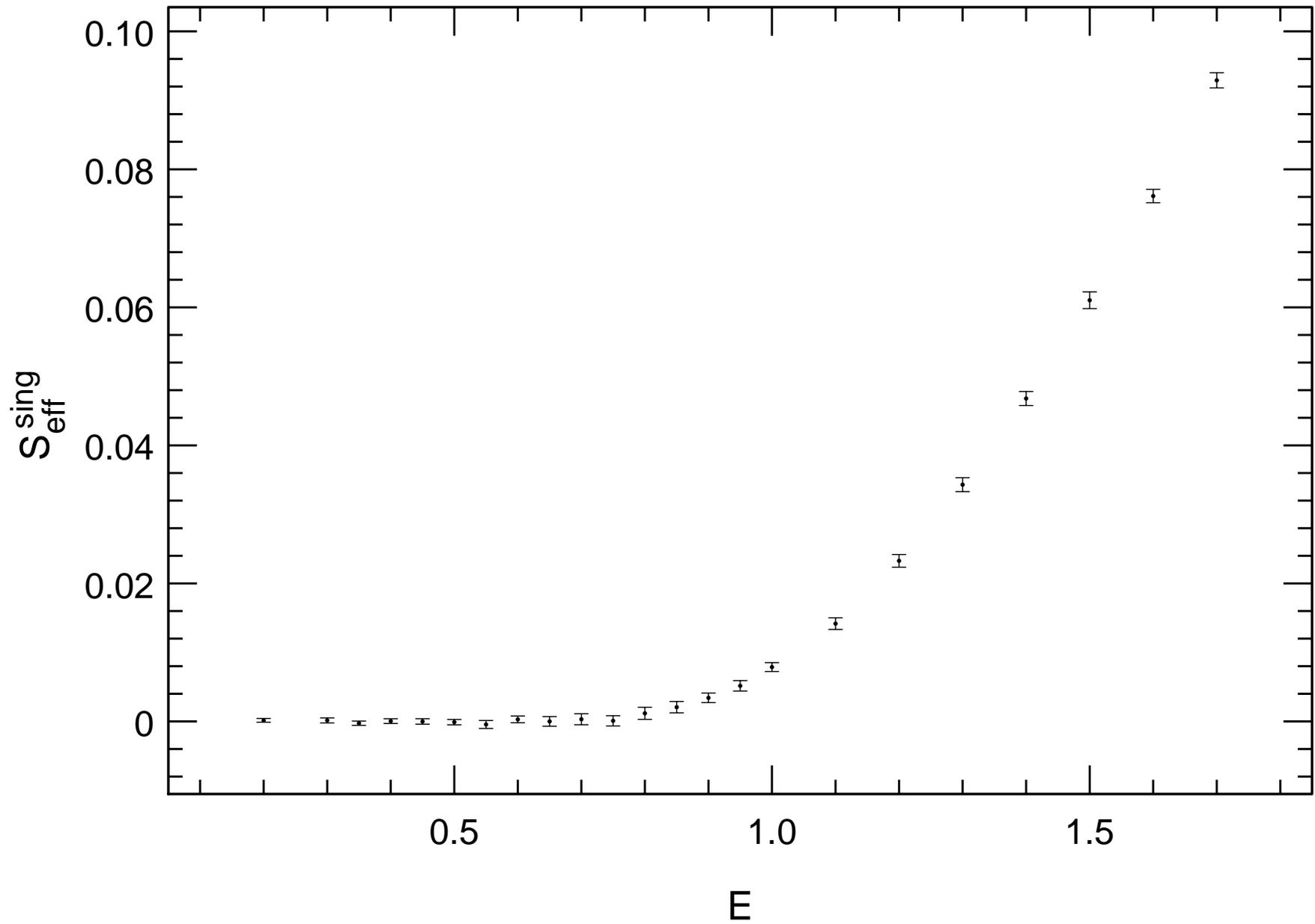
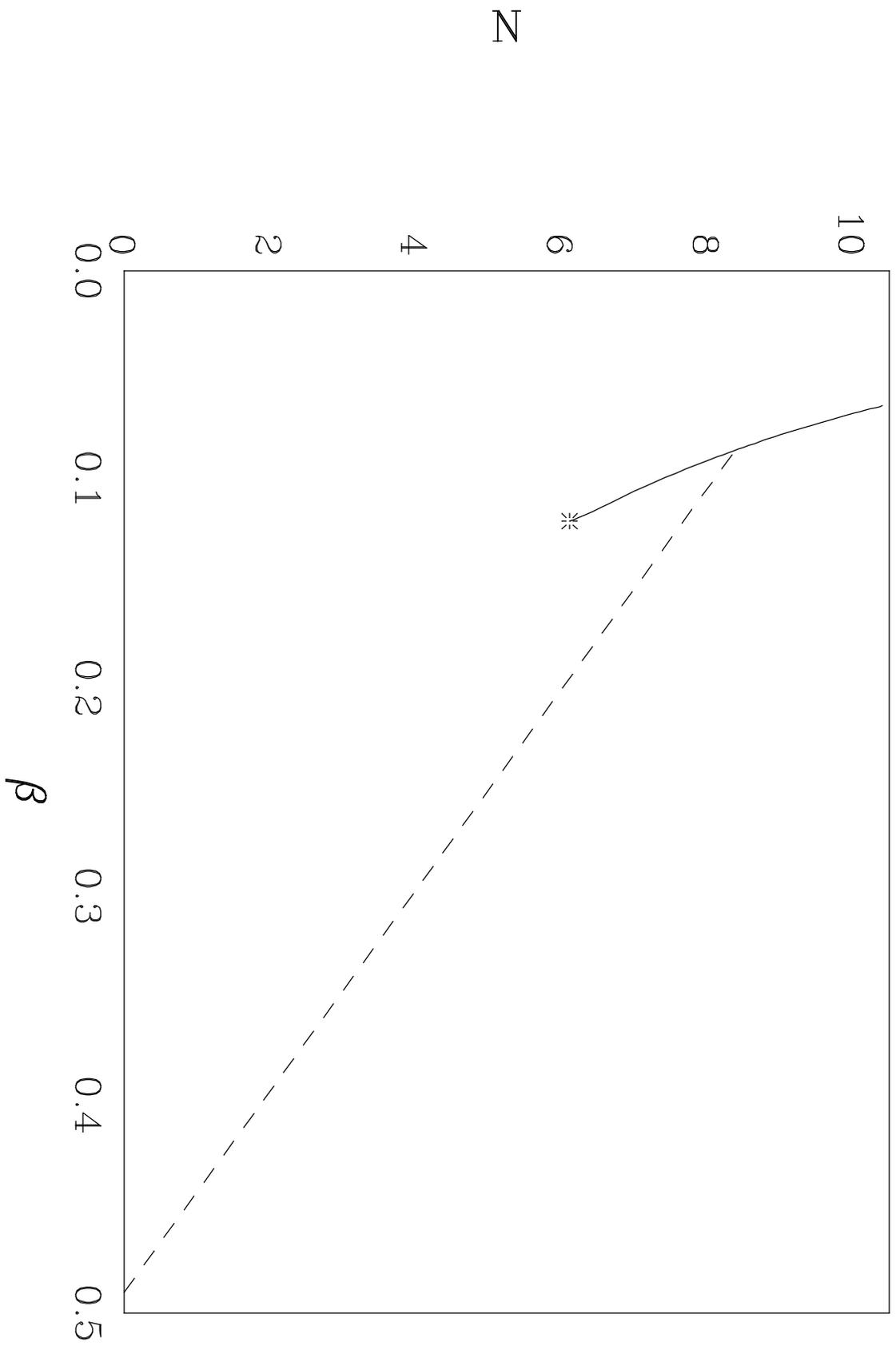
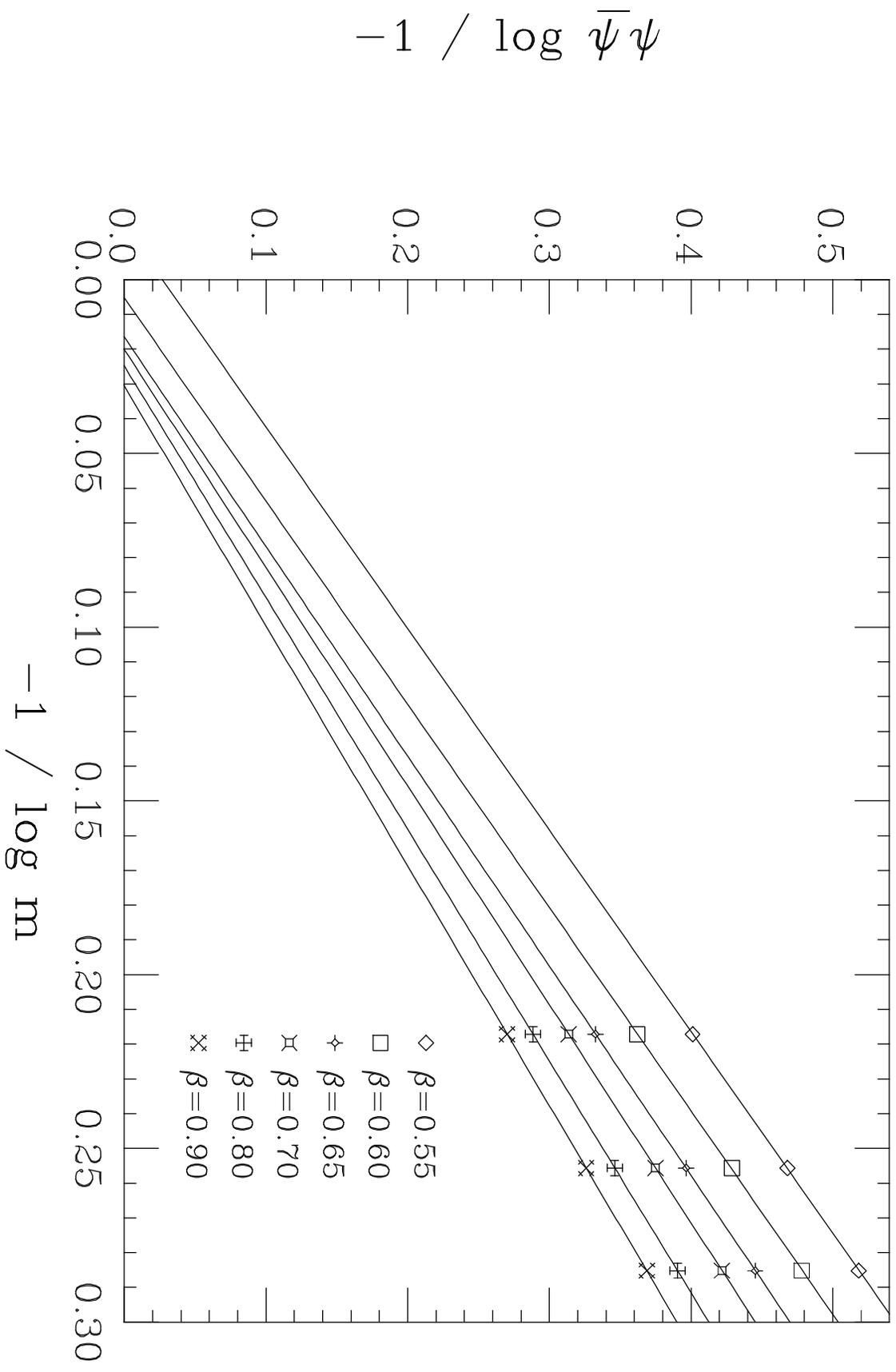
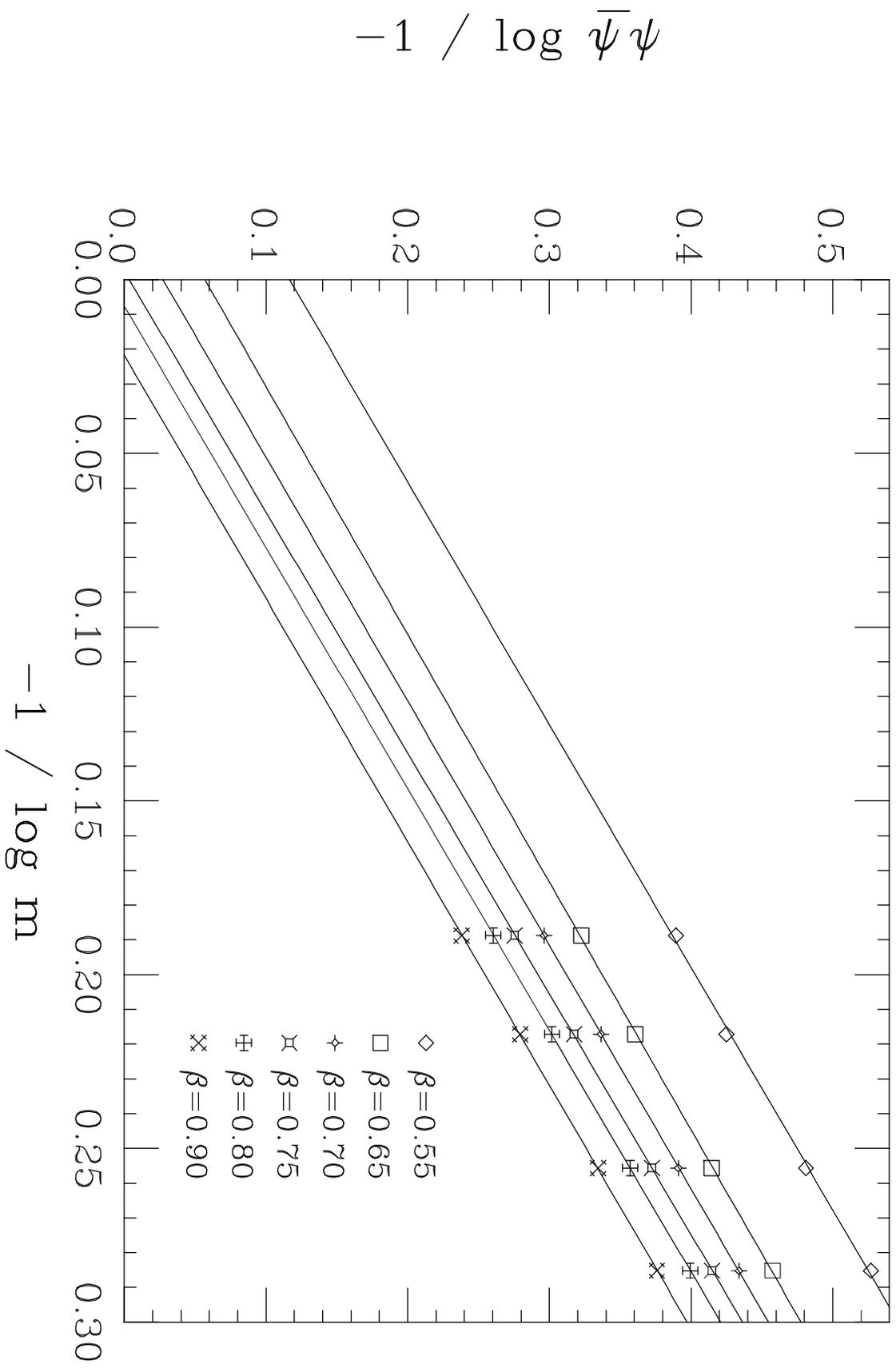
\end